# Raman Scattering by $sp^2$ Amorphous Carbons

E.F. Sheka[1*], Ye. A. Golubev[2], N.A. Popova[1]

[1] Peoples' Friendship University (RUDN University) of Russia, Miklukho-Maklay 6, 117198 Moscow, Russia
[2] Yushkin's Institute of Geology, Komi Science Center, Ural Branch of RAS, Pervomayskaya 54, 167982, Syktyvkar, Russia

*Corresponding author: E.F. Sheka

E-mail: sheka@icp.ac.ru

**Abstract.** The paper presents a cooperative consideration of Raman spectra of $sp^2$ amorphous carbons as well as the nature and type of their amorphicity. The latter was attributed to the amorphization of a new type named as enforced fragmentation. The fragments are stable graphenous molecules, which are the basic structural units (BSUs) of the solids, determining them as amorphics with molecular structure. Due to weak intermolecular interaction, BSUs, once aggregated, are the main defendants for IR absorption and Raman scattering of the solids, just justifying the consideration of them at molecular level. The standard G-D-2D pattern of Raman spectra of polycyclic aromatic hydrocarbons, $sp^2$ amorphous carbons, graphene and/or graphite crystal is attributed to extended honeycomb composition of carbon atoms and are suggested as manifestation of molecule-crystal dualism of graphenous materials. The molecular approximation, applied to the analysis of one-phonon spectra of the studied $sp^2$ ACs, makes it possible to trace a direct connection of the G-D spectra image as well as their broadband structure with a considerable dispersion of the C=C bond lengths within BSUs honeycomb structure, caused by the influence of chemical action, deformation, etc. This approximation, applied to the interpretation of two-phonon spectrum of graphenous molecules for the first time, reveals a particular role of electrical anharmonicity in the spectra formation and attributes this effect to a high degree of the electron density delocalization. A size-stimulated transition from molecular to quasi-particle phonon consideration of Raman spectra was experimentally traced, which allowed evaluation of a free path of optical phonons in graphene crystal.

> Like our own signatures, Raman spectra, when properly analyzed, can reveal much about the character as well as the identity of the molecular signatory
>
> Derek A. Long.

## 1. Introduction

Seemingly simple and even routine at first glance, the plot in the title of the article is actually far more complex. Difficulties concern both personages of the plot. Due to rapid development of graphenics, Raman scattering has become in recent years a massive usual method of testing a variety of graphenous materials. A well-known solid-state approach, based on applying the difference in intensities of the components of either G-D doublet or G-D-2D triplet of characteristic bands for the fixation of the objects structuring (see germinal works [1-8]), filled thousands of pages of publications. In the vast majority of cases, this approach is used entirely formally without analysis of what graphenous material under study is and what, in fact, its Raman spectrum is talking about. The purpose of this article is to show, by the example of Raman scattering by $sp^2$ amorphous carbons (ACs), that the abyssal content of the spectra obtained is closely related to the properties of the scattering object and contains important and unique information about it that goes beyond the formal G-D and/or G-D-2D considerations. To this end, we have studied under the same condition Raman spectra of seven elitist $sp^2$ACs of different origin and of the highest carbonization rank, comprehensively tested concerning their structure and chemical content, and two referent graphites of different structure. A comparative analysis of the obtained spectra, complemented with a detailed consideration of the object amorphicity, allowed finding a deep connection of the spectra patterns with the objects structural data and chemical content as well as pointing to a necessary discrimination of conditions when either molecular or solid-state approach are appropriate for the spectra interpretation.

The paper is composed as follows. Section 2 is devoted to the modern nomenclature of amorphous carbons. An overview of the Raman spectra of the studied samples and problems concerning their interpretation are presented in Section 3. Specification of the ACs amorphicity and solid-state approach of the spectra consideration are discussed in Section 4. General comments concerning vibrational spectroscopy of amorphous carbons as well as main concepts of molecular approach related to graphenous molecules are considered in Section 5. Sections 6 and 7 present the interpretation of the obtained Raman spectra in the frame work of molecular approximation, concerning one- and two-phonon fractures, respectively. Conclusion summarizes main essentials received.

## 2. Today's nomenclature of amorphous carbon

Carbon, one of the most versatile elements around, manifests itself in a wide variety of allotropic forms that exhibit a diverse range of properties. ACs (*amorphics* for simplicity) are widely common and form a big class of members, both natural and synthetic. Natural amorphics are products of the activity of natural laboratory during geological billion-million-year time. Exhausted geological examinations allowed suggesting a few classification schemes of carbon materials produced, one of which, the most popular, slightly changed with respect to original [9], is schematically shown in Fig. 1a. The scheme presents a continuous evolution of a pristine carbonaceous mass into ordered crystalline graphite thus exhibiting the main stream of carbon life in the Nature. The evolution is presented as increasing carbonization rank of intermediate products. As seen in the figure, a general picture can be split into two gloves, the left of which starts with plants and sediments of different kinds and proceeds through sapropels to brown coals and later to usual coals and anthracite. The endpoint on the way is graphite. As for the right glove, it covers carbonization of pristine gas and distillate oil and proceeds through petroleum and naphthoids to asphalts and then to kerites, anthraxolites, and shungites. As in the previous case, graphite is the endpoint. Certainly, the division is not exactly rigid due to which a mixture of the two fluxes, particularly, at early stages of carbonization, actually occurs. This scheme is related to $sp^2$ amorphous carbon that actually dominates in the Nature. $sp^3$ Amorphous carbon is not so largely distributed over the Earth due to which diamond-like natural ACs have become top issues of the carbon mineralogy [10, 11] only recently. Natural amorphic sampling for the current study concerns anthracite, anthraxolite and shungite.

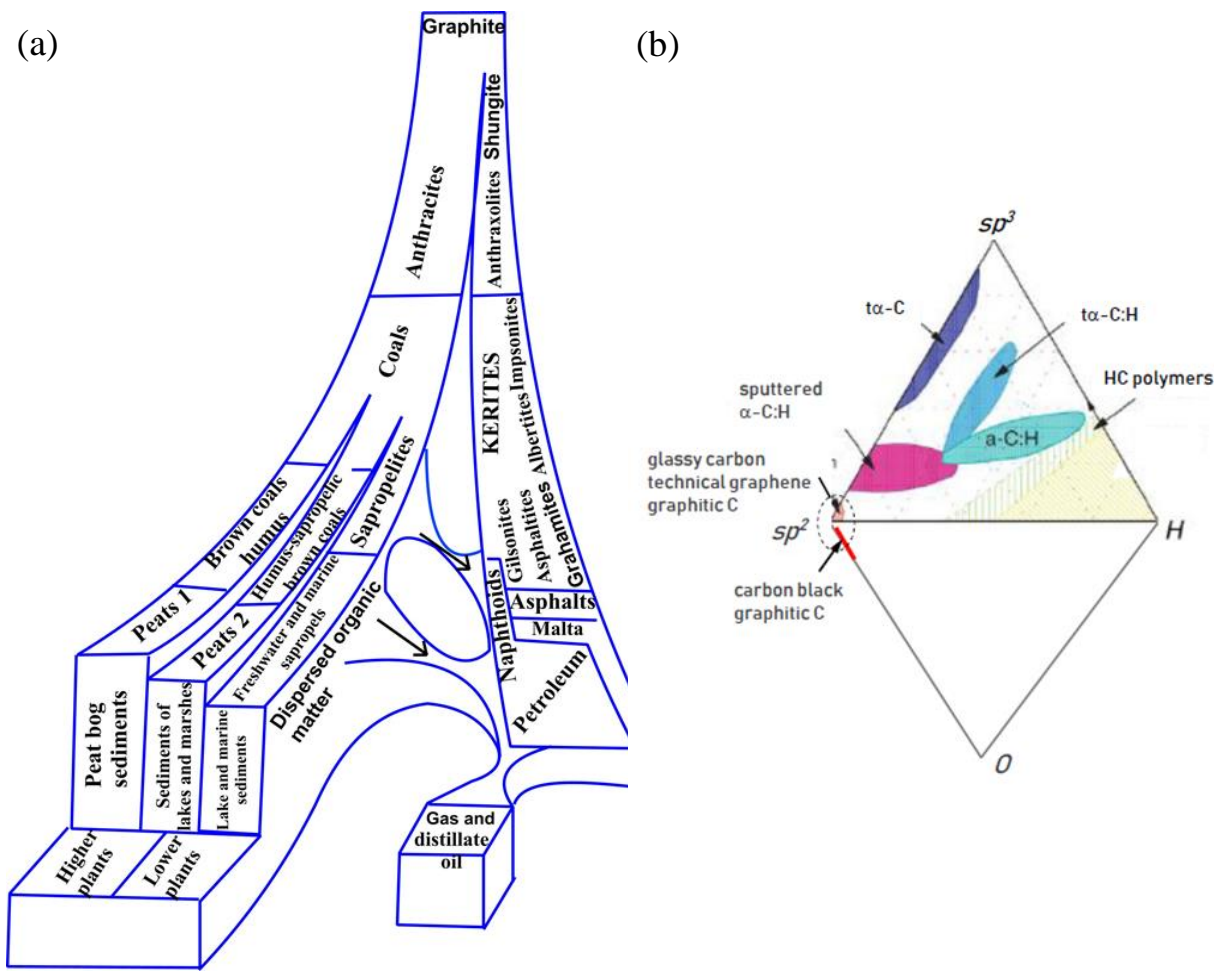

**Figure 1.** a. Carbon life path in the Nature: Amorphous carbons based on the Uspenskiy's classification [9]. b. Rhombic phase diagram of the amorphous carbon-hydrogen-oxygen system.

The family of synthetic amorphous carbons is quite large, covering species different by not only the carbonization rank, but also a combination of $sp^2$ and $sp^3$ components in the presence of hydrogen. Analogously to natural, synthetic amorphics were classified as well [12] and the relevant classification scheme is presented in Fig. 1b. Previously ternary, it is completed in the current study to a rhombic one to take into account oxygen as one more important ingredient, which is caused by large development of modern techniques of the ACs production. Comparing schemes presented in Figs. 1a and b, it becomes evident that those are related to fully different communities of substances. If natural amorphics belong to $sp^2$ carbon family and are rank-characterized with respect to the stage of their metamorphism and carbonization, synthetic amorphics are mainly characterized by $sp^3$-configured solid carbon. $sp^2$ Group of the synthetic species takes only small place, marked by oval, in the total family and their carbonization is the highest. In part, this circumstance is because in the middle of the previous century a great influence on material science was exerted by the active development of the physics of amorphous state, which was formed mainly in relation to monoatomic amorphous silicon. Expectedly, the attention was also paid to the upper and lower silicon neighbors in Mendeleev's periodic table, which stimulated the synthesis of $sp^3$ amorphous carbon and germanium similar to that of silicon. Because of the absence of $sp^3$ solid carbon in nature, special technologies to produce t$\alpha$-C, t$\alpha$-CH, and sputtered $sp^2/sp^3$ mixed $\alpha$-C:H products were developed. As for $sp^2$ synthetic amorphics, for a long time they were presented by multi-tonnage industrial production of glassy (covering graphitic, black, activated and other highly carbonized) carbon [13]. However, the graphene era called to life a new high-tech material – technical graphenes [14] which are the final product of either oxidation-reduction or oxidation-thermally-shocked exfoliation of nanosize graphite [15, 16]). Two new members of this community are on the way – that are laser-induced graphene (LIG) manufactured by multiple lasing on cloth, paper, and food [17] and extreme quality flash graphene (FG) [18], the most approached by ordering to graphite.

In the current study, our attention will be concentrated on $sp^2$ ACs only. Recent comparative studies [19-27] together with a large pool of individual data have shown that these amorphics form a particular class of solid materials. As occurred, the general architecture of both natural and synthetic species is common and can be characterized as multilevel and fractal [28, 29], while differing in details at each level. Conceptually, the AC structure of the first level is well similar presented by *basic structure units* (BSUs). The higher level structure depends on the BSUs size. Thus, in the case of natural amorphics, the second-level structure is arranged by nanosize-thickness stacks of nanosize BSUs subjected to weak intermolecular interaction. These stacks form globules - a structure of the third level with pores of tens of nm. Further aggregation of globules leads to the formation of nano-microsize agglomerates characterized by pores of the corresponding size. Figure 2 presents schematically the evolution of such amorphic structure from a single BSU to macroscopic powder using a set of BSU model structures of shungite carbon suggested in [26, 27]. Synthetic amorphics are characterized by a large dispersion of BSUs lateral size from units to tens and/over first hundreds of nanometers. At the low-limit end of the dispersion, the amorphic structure is similar to that of natural species described above. At the high-limit end, the BSUs size does not prevent from BSUs packing in nanosize-thickness stacks while these laterally extended stacks are further packed in a paper-like structure.

According to studies [14, 26, 27], BSUs of both natural and synthetic $sp^2$ ACs present graphenous molecules, which, in their turn, are elements of a graphene honeycomb structure, framed around the perimeter by heteroatoms and/or atomic groups, including hydrogen, oxygen, nitrogen, sulfur, halogens and other elements. The molecules can be described by statistically averaged chemical formula (such as $C_{66}O_4H_6$ in the case of shungite carbon in Fig. 3) that corresponds to the chemical content of the sample obtained experimentally. Size, shape, chemical content of BSUs as well as terminating groups and their disposition in the circumference greatly vary, due to which each AC, classified usually by origin, covers a large class of specially framed graphene molecules. Nevertheless, despite the complexity of the overall fractal structure of $sp^2$ ACs, precisely BSUs determine the solids properties and stipulate molecular-structural approach for their description. Firstly applied to vibrational spectra of the solids provided by INS and DRIFT spectroscopies [25, 27], results of which were successful, in the current paper the approach is extended to the Raman spectra.

To this end, specific $sp^2$ AC representatives of the highest-rank carbonization, well tested in relation to their structure and chemical composition, were selected [26, 27]. This set involved samples of two carbon blacks (CB632 and CB624), one chemically (Ak-rGO) and one thermal-shock exfoliated (TE-rGO) technical graphenes as well as shungite carbon (ShC), anthraxolite (AnthX), and anthracite (AnthC) (see the detailed description of samples in [27]). The set is complemented with two Botogol'sk graphites of the best quality [30] characterized by mono-

(mncr) and microcrystalline (μcr) structure. The structural and chemical data of the samples are summarized in Tables 1 and 2. The data obtained earlier are supplemented in this study by the results of the X-ray diffraction and EDS measurements for the two graphites.

**Table 1.** Structural parameters of amorphous carbons [1]

| Samples | d (Å) | $L_{CSR}^c$, nm | Number of BSU layers | $L_{CSR}^a$, nm | Ref |
|---|---|---|---|---|---|
| mncr Gr[2] | 3.35(X) | 105 | 313 | 550 | this work |
| μcr Gr[2] | 3.35(X) | 49 | 146 | 184 | this work |
| ShC | 3.47(N); 3.48(X) | 2,5(N); 2.0(X) | 7(N); 5-6(X) | 2.1(X) | [26] |
| AnthX | 3.47(N); 3.47(X) | 2.5(N); 1.9(X) | 7(N); 5-6(X) | 1.6(X) | [26] |
| AnthC (Donetsk) | 3.50(X) | 2.2(X) | 5-6(X) | 2.1(X) | [27] |
| CB632 | 3.57(N); 3.58(X) | 2.2(N); 1.6(X) | 6(N); 4-5(X) | 1.4(X) | [26] |
| CB624[3] | 3.40(N); 3.45(X) | 7.8(N); 6(X) | 23(N); 17(X) | 14.6 | [26] |
| Ak-rGO | 3.50(N) | 2.4 (N) | 7(N) | >20(N)[4] | [21] |
| TE-rGO | 3.36(N) | 2.9 (N) | 8(N) | >20(N)[4] | [22] |

[1] Notations (N) and (X) indicate data obtained by neutron and X-ray diffraction, respectively.
[2] The data are obtained by the treatment of (002) and (110) reflexes using Scherrer's relation $L_{CSR} = K·λ/B·cosΘ$. Here λ is the X-ray radiation wavelength (CuK$_α$) 0.154 nm, Θ is the position of the (110) ($L_{CSR}^a$) and (002) ($L_{CSR}^c$) peaks, B is the half-height width of the peak in 2Θ (rad)units, and constant K constitutes 0.9 and 1.84 for reflexes (002) and (110), respectively, in the approximation of a disk-shaped particles.
[3] X-ray data were corrected in the current study.
[4] $L_{CSR}^a$ = 20 nm means the low limit of the value accessible for the measurements performed.

## 3. An overview of Raman spectra of *sp²* amorphous carbons and problems to be solved

Despite the fact that the Raman spectra (RSs) of *sp²* ACs were recorded countless number of time (suffice to say that RS is an indispensable participant in testing each natural and synthetic graphenous material), a joint test of different-origin representatives of these materials has not yet been carried out. The current study is the first when nine products of such kind are exhibited simultaneously. Raman spectroscopy was carried out with a LabRam HR800 instrument (Horiba, Jobin Yvon) at room temperature. The system was equipped with an Olympus BX41 optical microscope and a Si-based CCD detector (1024 × 256 pixels). A 50× objective (working distance ~3 mm, numerical aperture 0.75) was used. Spectra were recorded in the 100–4000 cm$^{-1}$ range using a spectrometer grating of 600 g/mm, with a confocal hole size of 300 μm and a slit of 100 μm. As exciting radiation external Ar+ laser (514.5nm 1.2 mW) was used. Each spectrum was the result of three accumulations with a 10 s exposure. The spectra decomposition, if necessary, is provided with LabSpec 5.39 program using a pseudo-Voigt function (Gaussian-Lorentzian Sum) basis.

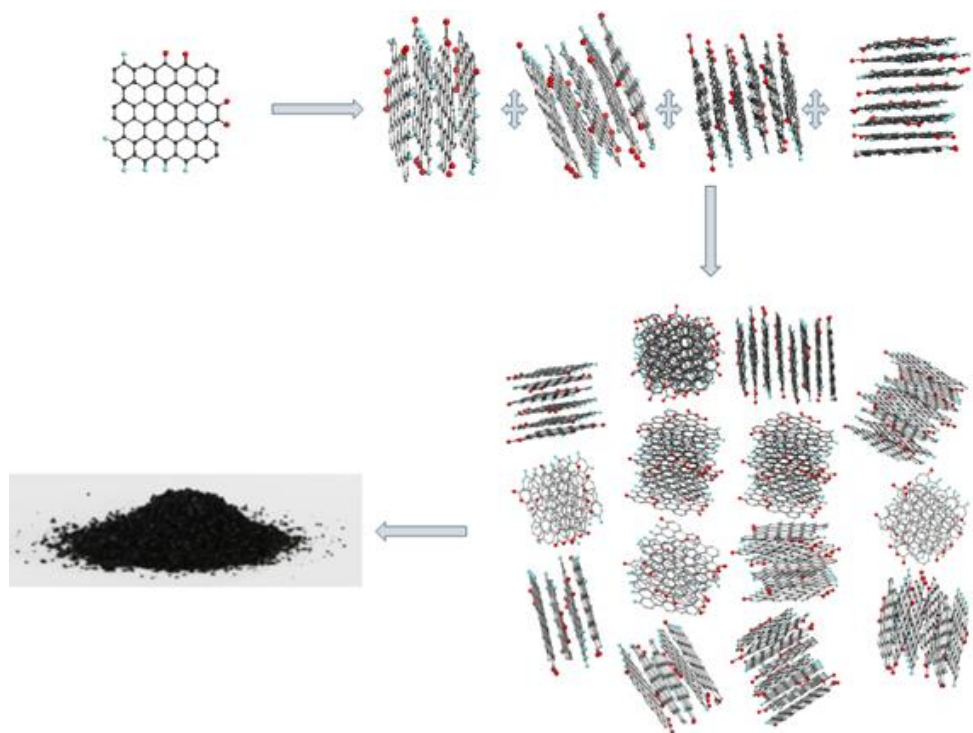

**Figure 2.** Schematic transformation from a single BSU molecule to powdered solid amorphous carbon via BSU stacks and globule(s). Planar view on a model globule with linear dimensions of ~6 nm. Dark gray, light blue, and red balls depict carbon, hydrogen, and oxygen atoms, respectively (see text).

**Table 2**. Chemical content of amorphous carbons

| Samples | Elemental analysis, wt% | | | | | Ref. | XPS analysis, at% | | | Ref. |
|---|---|---|---|---|---|---|---|---|---|---|
| | C | H | N | O | S | | C | O | Minor impurities | |
| mncr Gr[1] | 99.0 | - | - | 1.0 | - | This work | | | | |
| μcr Gr[1] | 98.9 | - | - | 1.1 | - | This work | | | | |
| ShC | 94.44 | 0.63 | 0.88 | 4.28 | 1.11 | [26] | 92.05 | 6.73 | **S** - 0.92; **Si** – 0.20; **N**-0.10 | [26] |
| AnthX | 94.01 | 1.11 | 0.86 | 2.66 | 1.36 | [26] | 92.83 | 6.00 | **S** - 0.85; **Si** – 0.25; **N**-0.07 | [26] |
| AnthC | 90.53 | 1.43 | 0.74 | 6.44 | 0.89 | [27] | 92.94 | 6.61 | **Cl** - 0.11 - **S**: 0.34 | [26] |
| TE-rGO | 84.51 | 1.0 | 0.01 | 13.5 | 1.0 | [27] | 86.77 | 10.91 | **F** - 077; **S** - 0.86; **Si** -0.70 | [27] |
| Ак-rGO | 89.67 | 0.96 | 0.01 | 8.98 | 0.39 | [27] | 94.57 | 5.28 | **S** - 0.16 | [27] |
| CB624 | 99.67 | 0.18 | 0 | 0.15 | - | [26] | 95.01 | 4.52 | **Si** – 0.46 | [26] |
| CB632 | 97.94 | 0.32 | 0.04 | 1.66 | 0.68 | [26] | 93.32 | 6.02 | **Si** – 0.66 | [26] |

[1] EDS measurements

Figure 3 presents a general view of RSs related to the studied solids. The spectra are collected in two columns related to natural (left) and synthetic (right) samples. All the spectra are of a characteristic view decorated with three sharp features. Paying tribute to the established terminology based on the DRRS approximation [1-8], we use G and D as well as 2D markers to designate bands of the obtained spectra related to one- phonon and two-phonon spectral regions, respectively.

Looking at this collection, we would like to start with the first features that concern fine-structured spectra located in the first row in Fig. 3. Among the latter, there are two lineaments, which require a particular attention. The former is related to the RSs of graphites while the latter – to the RS of amorphic CB624. As for graphites, a scrupulous analysis of the available RSs reveals that a single G-band-one-phonon spectrum is a rarity, once characteristic for "the best" graphites such as Madagascar flakes and Ticonderoga crystals [1], Ceylon graphites [31], Botogol'sk graphites [31, 32], and some others. In contrast, in the predominant majority of cases, researchers are dealing not with monocrystalline, but highly structured graphites. As known, natural graphitization occurs not in a vacuum, but in a certain environment. Consequently, according to the main laws of nanosize-object chemistry [33], the termination of dangling bonds in the circumference of individual graphene sheets stops the growing of the latter thus transforming monolithic graphite massive into micronanostructured one with a characteristic G-D doublet pattern of their RSs. Only particular circumstances such as high temperature and pressure could resist the graphitization stopping that is why large pieces of monocrystalline graphite are usually found at great depth in the Earth core, as is the case of Botogol'sk deposit. However, even under such conditions graphite rocks are not structurally homogeneous, consisting of microscopic monocrystalline blocks surrounded with micronanocrystalline graphite. Graphite spectra in Fig. 3 just exhibit these two mncr and μcr constituents. Expectedly, BSUs of the studied graphites markedly differ in lateral size (see $L_{CSR}^a$ data in Table 1), once terminated by oxygen, hydrogen, and other heteroatoms and/or groups, among which the weight content of oxygen constitutes ~ 1 wt% (see Table 2).

Similarity of the spectrum of amorphous CB624 and those of the discussed graphites convincingly evidences that not only in the depth of the Earth core, but also in industrial reactors producing carbon black a significant graphitization of amorphous carbon may occur. And if according the data listed in Table 1 the extent of the CB624 graphitization is modest enough, the RS willingly responds to it making the CB624 spectrum to be quite similar to the spectrum of μncr graphite. A detailed discussion of the peculiarity will be presented in Sections 6 and 7.

The spectra of natural ACs as well as technical graphenes and carbon black CB632 keep the same G-D-2D pattern, but are significantly broadened. The spectra pattern convincingly evidences the commonality of the governing motives of the solids structure, which are provided by the relevant BSUs, thus reliably representing the commonality of graphenous nature of their carbon skeletons. Therefore, it becomes evident that just extended honeycomb composition of carbon atoms is responsible for the G-D-2D pattern of the amorphic RSs. In the current case, the lower size of the composition constitutes ~1.4 nm. However, the same RSs pattern is characteristic for molecules of 2D polycyclic aromatic hydrocarbons (PAHs) [34, 35], which shortens the size criterion to ~0.5 nm. Thus, we are facing an amazing manifestation of molecule-crystal dualism of graphene [36], which reveals itself in the form of a standard G-D-2D spectrum shape in the entire scale of the size range covering a particular chain of $sp^2$ carbonaceous species from 2D PAHs to graphite via ACs.

Intuitively, precisely this not yet conscious, but existing dualistic nature of the RSs peculiarity led the foundation of two approaches suggested for its explanation, namely, crystal and molecular ones. The former, known as DRRS [1-8], is widely explored with the researcher community and forms grounds for the mainstream of the interpretation of the RSs of graphenous materials. The latter, developed by an Italian team [37-47] has been so far outside of the mainstream. However, in the light of the dualistic peculiarities of the spectra discussed above, it deserves careful consideration. In what follows, both approximations will be analyzed from the viewpoint of their ability to explain the Raman scattering from graphenous materials within the whole size scale of the latter, mainly addressed to $sp^2$ ACs. A clear presentation of molecular-structural and solid-structural visions of $sp^2$ ACs is needed to make this analysis the most comprehensive. The former was discussed in the previous section. The latter is presented in the next section.

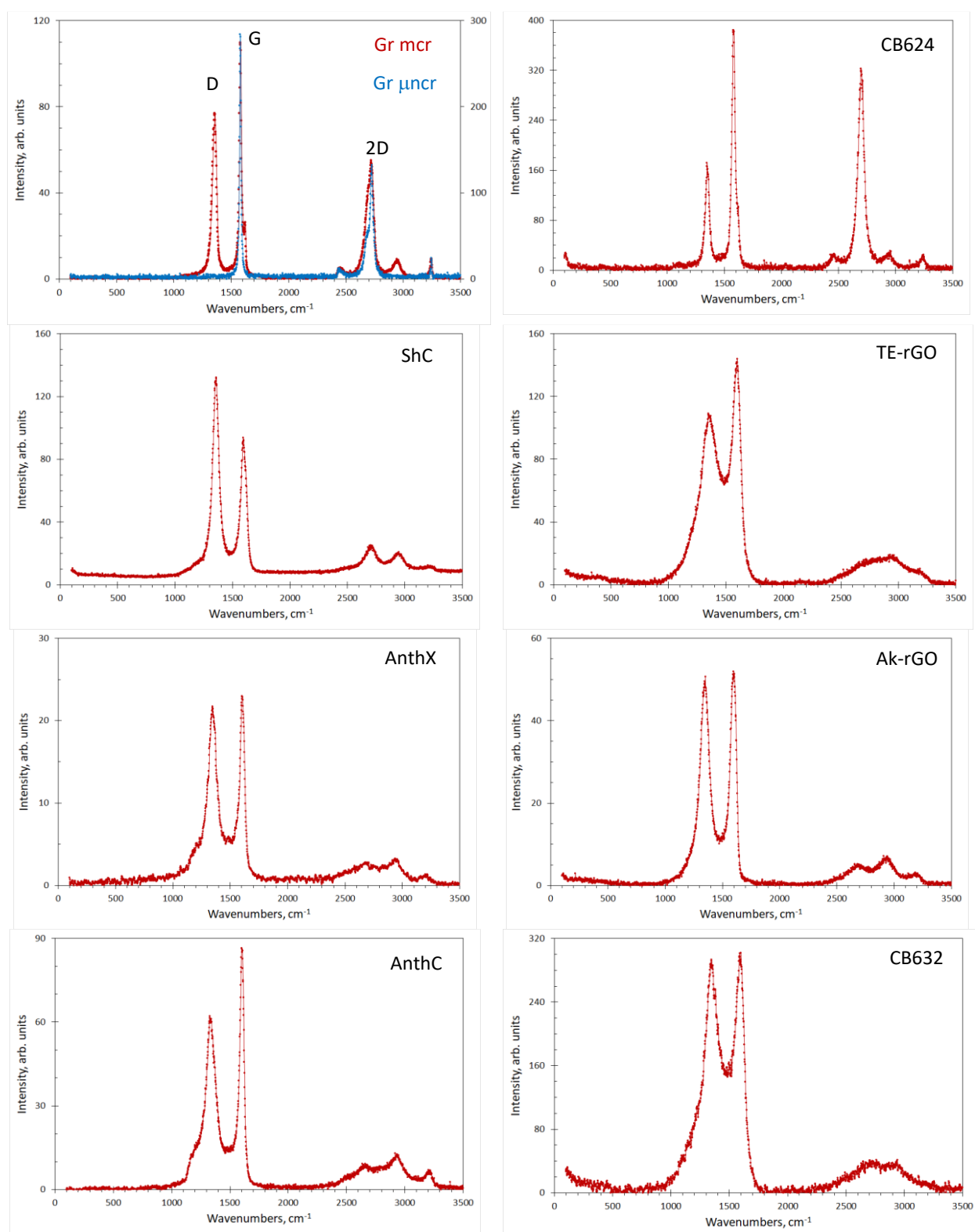

**Figure 3**. A general view of room-temperature Raman spectra of shungite carbon (ShC), anthraxolite (AnthX), anthracite (AnthC), technical graphene TE-rGO [20] and Ak-rGO [19], Merk-Sigma-Aldrich carbon blacks 699632 (CB632) and 699624 (CB624), and mono (mcr) - and micronanocstructured(µncr) graphites. respectively.

## 4. The amorphicity of $sp^2$ solid carbon

Despite $sp^2$ ACs have been the object of study and practical use for hundreds of years, until now they have not been considered from the general concept of the solid-state physics. The amorphicity of solids was widely studied and the main concept is discussed in a monograph [48]. The first issue of the concept concerns a considerable degree of amorphous solid ordering despite the lack of periodicity and/or translational symmetry. The ordering is subdivided into short-range (local) and medium range ones, the boundary between which passes around a few nm. The second issue is related to a direct connection of the solids properties and their local ordering due to which establishing a local structure has always been the main goal of studying amorphous solids. In the case of monoatomic Si and Ge, it was found that tetrahedrally bonded atoms form the short-range order of the solids. The two chemical entities together with carbon form the common tetrel family of Mendelev's periodical table and can be subjected to a comparative analysis based on the phase diagram shown in Fig. 1b. Evidently, the data related to Si and Ge are concentrated near $sp^3$ corner indicating the similarity of these materials with t$\alpha$-C phase only. The same is related to Si-H and Ge-H, on the one hand, and t$\alpha$-C:H, on the other. The remaining part of the C-diagram is connected with the presence of $sp^2$-configured amorphics while there are neither silicious nor germanious analogues of the species. We believe that the absence of $sp^2$-configured amorphous phase of solid Si and Ge as well as 'aromatic' families of the species chemical compounds and freestanding honeycomb monolayer silicene and germanene is due to a fundamental reason connected with an extreme radicalization of their elongated double covalent bonds [49, 50]. Drastic difference of carbon materials from those of tetrels, caused by the feature, is considered in details in monograph [36]. Accordingly, from the standpoint of the general concept of amorphicity, monoatomic solid carbon has the unique ability to form amorphous (as well as crystalline) states of two types, characterized by fundamentally different short-range orders presented by either tetrahedral groups of bonded atoms or a honeycomb network of benzenoid units.

As mentioned earlier, extensive studies of a set of $sp^2$ ACs [14, 19-27] allows convincingly concluding that their short-range order is provided with framed graphenous molecules. The framing of molecules plays a decisive role, ensuring the formation and stability of short-range order, on the one hand, and preventing crystallization, on the other [28, 33], thus attributing the origination of $sp^2$ ACs to the *reaction amorphization* [48]. The medium-range order of the amorphics reliably follows from the fractal porous structure evidently observed experimentally [29, 51] and is presented by nanosize BSU stacks, composed in either globules (natural amorphics [28]) or paper-like sets of stacks (technical graphenes [21, 22]). A model structure of a globule, shown in Fig. 2, corresponds to shungite carbon. Molecule $C_{66}O_4H_6$ constitutes one of possible models of the ShC BSUs, related to structural and chemical data of the species listed in Tables 1 and 2. Composed into four-, five-, and six-layer stacks, the molecules create a visible picture of the medium-range order of the species. Analogously, medium-range-order globules of anthraxolite and anthracite can by visualized using the relevant BSU models suggested in [27]. Apparently, similar medium-range configuration can be suggested for industrial carbon black

CB632. Bigger lateral dimensions of BSUs of technical graphenes Ak-rGO and TE-rGO as well as carbon black CB624 make to think about laterally extended sets of several-layer stacks based on molecular BSUs similar to those presented in [27].

Concluding this brief discussion of the general view of the $sp^2$ carbon amorphicity, we should dwell on one more distinctive feature. Evidently, the molecular nature of BSUs makes it possible to reliably attribute $sp^2$ ACs to amorphics with a molecular structure. However, in contrast to the typical representatives of this class of solids, which are based on molecules of a stable standard structure [52, 53], BSUs of $sp^2$ ACs are not standardized and the data shown in Tables 1 and 2 represent only statistically averaged values. The transition from amorphous to ordered structure also occurs in different ways. In ordinary amorphics, this transition occurs as a gradual ordering of the positions of the constituent molecules with respect to each other via a sequence of mesomorphic transformation thus restoring translational periodicity (see details in [53]). In contrast, the crystallization, better to say, graphitization of $sp^2$ ACs occurs due to increasing the BSU size, as is shown schematically in Fig. 4. To trace processing, we have fixed the atomic C:O:H relative content by the data related to shungite carbon and have simulated the intermediate transition by growing the carbon skeleton. Certainly, the size-increasing of individual BSUs is accompanied with bridging of both BSUs themselves and stacks of them. For many years, this bridging was considered as a priority process of sequential ordering of the primary block-mosaic structure of natural ACs [54-57]. Obviously, the priority of one of the two processes evidently depends on the environment. Nevertheless, not depending on whichever process dominates in practice, we should come to a conclusion that $sp^2$ amorphous solid state can not be attributed to any of the known types of disorder characteristic for monoatomic solids and including topological no long-range one leading to irregular lattice as well as to spin, substitutional, and vibrational disorder of regular lattice [48]. What is observed empirically in the case of ACs, in terms of solid-state physics can be attributed to *enforced fragmentation* of ideally ordered graphite and/or graphene. Obviously, two types of fragmentation, from the top and bottom ones, can occur.

There might be various reasons for fragmentation, including mechanical impact, chemical reaction, temperature shock, exposure to hard radiation, etc., which can be easily traced by the history of the production of the ACs samples selected, in particular, for the current study [26, 27]. The fragments ensure the short-range order of the solids and are characterized by large variety with respect to not only different classes of ACs provided by different history and/or technology of their production, but the same class as well. The variety concerns size, shape, variation of chemical content, and, which is the most important, distribution of heteroatoms in the fragment framing area at fixed atomic percentage in average. Thus, models presented in Fig. 4, are only 'one snapshot' of communities related to possible permutations of hydrogen and oxygen atoms in the framing area that have no number. Therefore, the fragment structure of $sp^2$ ACs is highly inhomogeneous just attributing the solids to inhomogeneously amorphous structures.

Another distinctive feature of $sp^2$ ACs lies in the radical nature of their BSU fragments. As shown [26, 27], BSUs of ~2 nm in size are molecular radicals whose chemical activity is concentrated at non-terminated edge carbon atoms (these atoms are clearly seen on the top of Fig. 4) [27]. However, as can be seen from the figure, increasing the molecule size leads to decreasing the fraction of edge atoms due to which, at a fixed atomic composition C: O: H, the number of non-terminated atoms and the degree of the associated radicalization decrease. Thus, it is possible to think that weight content of oxygen in graphites in Table 2 corresponds to a complete termination of the relevant BSUs. Simultaneously, the presence of oxygen in the best graphites itself (the finding should be typical for all graphites [26]) evidences the limited size of BSUs as well as the unavoidable termination of the latter in the circumference. As for the weakening of the radicalization of BSU as its size grows, the issue, in particular, explains the well-known empirical fact that enforced nanostructuring of $sp^2$ ACs significantly enhances the yield of the reaction when the amorphics are used as carbocatalysts [58, 59]. With this idea of the main distinguishing features of the structure of $sp^2$ ACs, we now turn to Raman scattering by these solids.

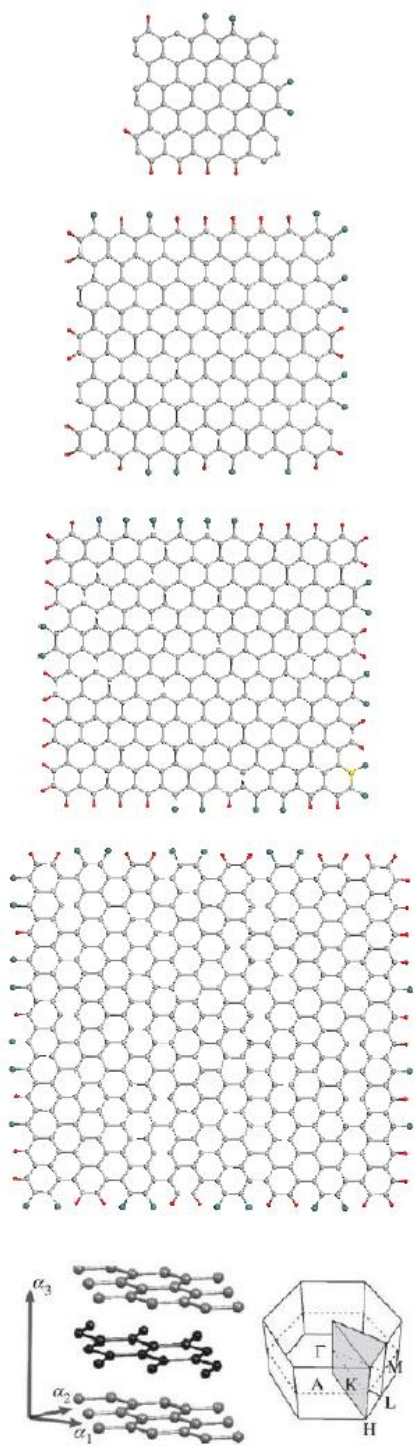

**Figure 4.** BSU models for sequential graphitization of amorphous carbon like shungite carbon. Carbon skeletons (n, n) NGr are wrightangle graphene sheets with *n* benzoid units along armchair and zigzag edges. Top down: (5, 5)NGr, (9, 9)NGr, (11, 11)NGr, (15, 12)NGr. Lateral dimensions are 1.2, 2.2, 2.7, and 3.3 nm, respectively. Light gray, red and blue balls depict carbon, hydrogen, and oxygen atoms, respectively. The chemical composition of the species is subordinated to the chemical content in Table 2. *Bottom*. Graphite lattice and Brillouin zone. *a*1, *a*2 and *a*3 mark the unit cell of graphite. The irreducible domain is spanned by the Γ–M–K–Γ triangle within the plane. Adapted from [4].

## 5. Briefly about grounds of Raman scattering of amorphous carbons

### 5.1. General concepts of crystal approach related to graphenous solids

For vibrational spectroscopy to become a valuable short-range structural probe, BSUs must be capable of being vibrationally excited independently from the surrounding amorphous matrix [48]. In the case of the studied $sp^2$ ACs, the requirement is fully met due to the presence of the BSU framing areas, on the one hand, and the evident vibrational decoupling of the molecules weakly interacting with each other, on the other. The next issue concerns the spectra internal content. As known, vibrational modes are IR-active if there is a change in dipole moment associated with vibrations. Covalent homopolar bonds of molecules do not normally possess a static dipole moment due to which their activity arises from a dynamic mechanism in which the charge is transferred from extended to compressed bonds [60]. Evidently, the effect depends on the charge value. Homopolar C=C bonds with nil static moments constitute the main contribution of structure elements of BSUs molecules of the studied amorphics. As for dynamic mechanism, its effect depends on the charge value, which for a predominant majority of carbon atoms is small. Consequently, vibrational modes related to carbon atoms are expected to be practically IR-inactive. The feature is actually characteristic for DRIFT spectra of the studied amorphics [27], intensity of which was provided by heteroatoms located in the framing area of the molecules with the strongest preference of C-H bonds characterized by the largest static moments. Therefore, IR spectroscopy of $sp^2$ ACs is a mean to study heteroatom framing of their BSUs. In contrast, the polarizability is quite favorable for homopolar bands and because these bonds dominate in the BSU structure Raman scattering becomes one of the main techniques to study the structure of carbon cores of graphenous materials.

Being structurally sensitive, Raman scattering is widely used in the study of amorphous solids [61, 62] and this particular method has been the main support for the classification of amorphics given above [48]. The latter was most clearly manifested in the study of monoatomic semiconductors αSi and αGe (see a profound review [61]). It was shown that in the majority of cases (the matter concerns different ways of the solids production), the amorphics were to be attributed to topological ones. Because of the lack of translational periodicity, the parent crystalline phases deprave their plain wave character since quasi-momentum $\boldsymbol{q}$ losses its meaning due to which $\sum_i \omega_i (q \cong 0)$ fine-structured spectra of crystals transform into G(ω) broadband spectra of amorphics exhibiting the density of phonon states [62, 63]. The effect in this case is especially striking due to the large bandwidth of both acoustic and optical phonons. The same character of the amorphicity was detected for amorphics with molecular structure, the only difference being that the spectra broadening under amorphization concern only acoustic phonons due to small bandwidth of optical phonons originated from vibrational modes of constituent molecules [53]. As for monoatomic amorphics, their disorder can be stimulated by substitutional disorder of regular lattice as well. Under this condition, quasimomentum $\boldsymbol{q}$ keeps its meaning to a great extent even if particular effects arise due to the presence of defects [61]. One of such effects is known as double resonant Raman scattering (DRRS) based on rigid relations between energies and quasimomenta of incident and scattered photons, on the one hand, and electron-hole pair and phonons, on the other. The phonon appearance is provided with the elastic scattering of electrons on defects [1, 3, 4]. This effect is most significant in the case of a $\boldsymbol{q}$-linear character of the electronic and phonon spectra of the crystal. Naturally, that graphite and graphene were the main objects to be considered from the standpoint of the DRRS approach.

Thus, according to fundamentals of the solid-state physics of amorphous materials [48], this issue could be expected if both the crystal and its amorphous mesomorph are either monoatomically (αSi and αGe) [61] or molecularly (say, MBBA or EBBA) [53] composed systems.

To make it applicable to *sp²*configured graphite crystal, the latter, despite its hexagonal honeycomb structure, was considered as a monoatomically arranged system with four atoms per unit cell [1]. Each atom is chemically bound with three other atoms in one layer and weakly interacts with the fourth atom from the second layer thus mimicking tetrahedral groups of bonded atoms related to the $sp^3$ configuration. The crystal phonon spectrum consists of nine optical phonon branches, peculiarities of whose dispersion in the Brillouin zone lead to sparks attributed to G and D bands. However, the amorphization of $sp^2$ solid carbon drastically changes the short-range order substituting the above tetrahedral groups with graphenous molecule of $\mathcal{N}$ atoms, which is accompanied with vibrational spectrum involving $3\mathcal{N}-6$ vibrational modes [27]. Despite the fact, almost from the first steps, a discussion of the DRRS, which is valid for an object with a preserved regular lattice, slightly disturbed by the presence of defects, was turned to real samples of $sp^2$ ACs, the actual disordering of which, is infinitely far from this model. Nevertheless, this approach, in practice transformed into G-D and/or G-D-2D ones *ex cathedra,* dominates at structural characterization of these solids. This paradoxical situation is caused, on the one hand, by the surprising simplicity and sameness of the experimental Raman spectra of both $sp^2$ ACs and micronanostructured graphites discussed in Section 3. On the other hand, it is favored by the unique possibility of fitting theoretical parameters based on the approximation of the linearity of the energy spectra of electrons and phonons that determines the intensity of the effect. Apparently, physicists who were the first to adopt the practical use of RS for structural testing of graphenous materials were attracted by the simplicity of practical recommendations proposed by the DRRS approximation not going into the depth of the physical grounds underlying them. The greatest demand among these recommendations is the use of the ratio of the intensities of the D and G bands, $I_D/I_G$, the half-width of the D bands, $\Delta\omega_D$, etc. Although the direct dependence of these parameters on the size and shape of graphenous fragments remains controversial, and in some cases absent, the number of publications including Raman testing of structural features of graphenous materials in the form described above is constantly growing (see [32,64-66] for a few).

### 5.2. General concepts of molecular approach related to graphenous molecules

The molecular approach was developed in the opposite direction in order to explain the features of graphite RSs by those obtained from Raman scattering by graphenous molecules, which lead to a crystal spectrum in the limit of infinite size of a molecule [37-47]. The approach starts from the consideration of the object vibration spectrum. Evidently, this spectrum of any graphenous molecule is multitudinous and multimode due to which a certain simplification is needed to make the spectrum discussable. In the previous study [27] we suggested the spectrum of benzene molecule to lay the foundation of the simplified consideration of the vibrational spectra of $sp^2$ ACs BSUs. The list of benzene vibrational modes and their assignment are given in Table 3. According to symmetry rules, modes 1-10 are active in Raman scattering while modes 11-20 – in IR absorption. Evidently, any lowering of the molecule symmetry violates the double degeneracy and mixes the modes. Anyway, even with these limitations both Raman and IR spectra of benzene molecule should be quite rich. However, in practice the spectra of gaseous benzene are very simple and consist of small number of selected modes. Among the latter only two of them, $\nu_1$ and $\nu_7$, form the pattern of the observed RS. When the number of benzene rings increases, considerable changing of the spectra patterns occurs. As for linear chains of benzene rings, the governing role in the spectra, starting from naphthalene, goes to mode $\nu_8$ [66]. The corresponding band is surrounded by satellites due to the violation of the $D_{6h}$ symmetry of benzene and the removal of the degeneracy of the initial mode, first weak in naphthalene and anthracene, and then comparable in intensity in tetracene and pentacene. At the same time, a scattering of noticeable intensity in the region of modes $\nu_8$ and $\nu_{19}$ arises in the spectrum of the last two molecules. Therefore, the transformation of RS of benzene when going to pentacene consists in enhancing the role of C=C stretching vibrations originated mainly from $e_{2g}$ and $e_{2u}$ as well as $\nu_8$ and $\nu_{19}$ benzene modes. The shape of the spectrum in this region is still rather complex.

**Table 3.** Vibrational modes of benzene molecule (adapted from [67] )

| # | Symmetry | Frequency | Description |
|---|---|---|---|
| 1 | $a_{1g}$ | 993 cm$^{-1}$ | Breathing |
| 2 | | 3073 | C–H stretching in-phase |
| 3 | $a_{2g}$ | 1350 | C–H in-plane bend. in-phase |
| 4 | $b_{2g}$ | 707 | C–C–C puckering |
| 5 | | 990 | C–H out-of-plane trigonal |
| 6 | $e_{2g}$ | 606 | C–C–C in-plane bending |
| 7 | | 3056 | C–H stretching |
| 8 | | 1599 | C–C stretching |
| 9 | | 1178 | C–H in-plane bending |
| 10 | $e_{1g}$ | 846 | C–H out-of-plane (C$_6$ libration) |
| 11 | $a_{2u}$ | 673 | C–H out-of-plane in-phase |
| 12 | $b_{1u}$ | 1010 | C–C–C trigonal bending |
| 13 | | 3057 | C–H trigonal stretching |
| 14 | $b_{2u}$ | 1309 | C–C stretching (Kekulé) |
| 15 | | 1146 | C–H in-plane trigonal bending |
| 16 | $e_{2u}$ | 404 | C–C–C out-of-plane bending |
| 17 | | 967 | C–H out-of-plane |
| 18 | $e_{1u}$ | 1037 | C–H in-plane bending |
| 19 | | 1482 | C–C stretching |
| 20 | | 3064 | C–H stretching |

Another feature is observed when going from polyacenes to two-dimensional (2D) π conjugated planar structures of PAHs consisting of spatially extended composition of benzenoid rings. The first objects were PAHs C$_{78}$H$_{32}$ [34] and C$_{96}$H$_{30}$ [35] synthesized by Prof. Müllen's team. Raman spectra of both molecules [37] are not much distinguishable from the spectra of ACs presented in Fig. 3. Later on the PAH set was enlarged including molecules of different shape, symmetry, and carbon content from C$_{24}$H$_{12}$ to C$_{114}$H$_{30}$ [38, 39], whose RSs had a characteristic G-D-2D pattern. As turned out, the characterization concerns the general pattern of the spectra, while $I_D/I_G$, and $\Delta\omega_D$ parameters were quite individual. Thus, it was experimentally shown that the packing of benzoid cycles into a 2D planar structure leads to the characteristic G-D-2D appearance of the RS, which does not depend on the size and symmetry of the PAH molecules.

A profound theoretical analysis performed by the Italian spectroscopists [37-47] allowed both to reveal the reasons for the discovered uniqueness of PAHs RSs, and to establish their intimate connection with the spectra of graphite and/or graphene. This analysis was based on a thorough study of the dependence of the polarizability tensor of the molecules, which determines RS intensity, on the dynamic characteristics of the molecules under conditions of a multimode structure of the vibrational spectrum and various point symmetries. It was found that the main contribution to the intensity of the spectra is made by C=C stretchings, due to which the observed G–D spectra are characteristic spectra of the network of C=C bonds mainly. As for the stretchings themselves, the modes that determine G band originate from the $e_{2g}$ vibration of benzene, while the modes responsible for D band come from the $e_{1u}$ vibration of the molecule (see Table 3). The modes individuality is caused by peculiarities of their vibrational forms. As convincingly shown (see Fig. 5), the vibrations of G band correspond to simultaneous in-plane stretchings of all C=C bonds, while those related to D band concern both stretching and contraction of these bonds when carbon atoms move normally to them just imitating benzenoid ring breathing. The motion of carbon atom has a collective character, for which planar packing of benzenoid units is obviously highly preferable. In contrast, as seen in Fig. 5a, the vibrational forms of both $e_{2g}$ and $e_{1u}$ modes of benzene are local and different from collective forms of polybenzenoid molecules.

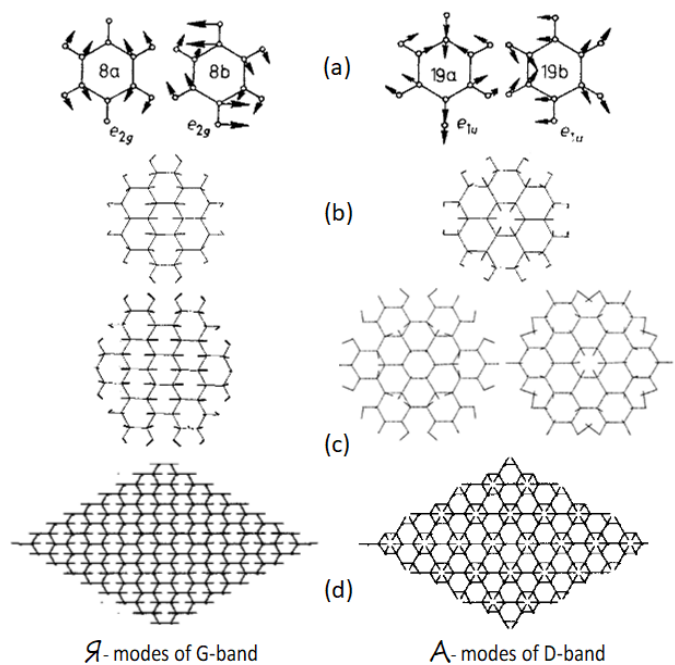

**Figure 5**. Vibrational forms of selected normal vibrations: (a) benzene; (b) coronene; (c) hexabenzocoronene; (d) a perfect 2D lattice of graphene. Adapted from [39] and [67].

Polarization of molecules is highly sensitive to the vibrational form. Actually, the quantity is generally described as [67]

$$\alpha_t = \alpha + \sum_i \left(\frac{\partial \alpha}{\partial Q_i}\right)_0 Q_i + \frac{1}{2!}\sum_{ik} \left(\frac{\partial^2 \alpha}{\partial Q_i \partial Q_k}\right)_0 Q_i Q_k + \cdots \quad (1)$$

where α is the polarizability in equilibrium position, Q's are the co-ordinates of the individual normal vibrations, sets $\{Q_i\}$ and $\{Q_k\}$ present vibrational forms of the $t^{th}$ vibration, and the subscripts $_0$ of the differentials refer to the equilibrium position. The third and subsequent members of the power series represent the electrical anharmonicity. The intensity of one- and two-phonon Raman scattering is governed by the second and third terms, respectively. As seen from the equation, the vibrational forms are directly involved into the intensity determination, differently for the Raman scattering of the first and second order.

Detailed consideration of the PAHs polarizability performed in the extensive study [37-47] showed that parallel-to-bond vibrational forms, attributed by the authors to the type Я [39]), promote a steady intense one-phonon Raman signal G in all the studied molecules. The feature does not depend on the molecules symmetry and shape and is caused by the $\frac{\partial \alpha}{\partial Q_i}$ derivatives, which all are positive. In contrast, normal-to-bond vibrational forms of type A [39] promote in this case both positive and negative $\frac{\partial \alpha}{\partial Q_i}$ derivatives due to which the intensity of D band is tightly connected with the "quality" of C=C bonds which influences the normal vibration forms. So, if the bonds are identical, the contribution of A modes into the one-phonon signal is nil. In the opposite case, the signal is not nil and is the bigger, the bigger the difference between the bonds

[40]. This conclusion has been approved on a number of PAH molecules, whose RSs were obtained and analyzed [37-40, 42, 44, 46]. Coming back to benzene, it becomes evident that the lack of collective character of the vibrational forms of $e_{2g}$ and $e_{1u}$ modes (see Figs 5a) does not favor the modes appearance in the molecule RS. Moreover, due to identity of C=C bonds and symmetry rule, D band must not be observed at all while G band of small intensity might be seen that is the real case.

From the very beginning of the study, Italian researchers have borne in mind RS of graphite and/or graphene crystals as a final point of the PHAs evolution by size. They extended their consideration over graphene crystal and have found that $e_{2g}$ mode at $\Gamma$ point of the first Brillouin zone is typical Я-mode while $A'_1$ mode at $K$ point clearly reveals A character (Fig. 5d). The findings undoubtedly evidence a peculiar collective character of the graphene phonons caused by benzenoid-hexagon structure. Therefore, until the honeycomb packing of benzenoid units is not broken, the graphenous molecules and supramolecules are characterized by the G-D patterned Raman spectra. Only a complete destruction of the sheet leads to this feature loosing, which was really observed for graphene after its extremely strong bombardment by Ar[+] ions [69].

Figure 5, in general, is deeply consonant with Fig. 4 and can be considered as a Raman-spectrum image of sequential graphitization of amorphous carbon. Bearing this in mind, let us analyze RSs of the studied ACs in detail. To this end, consider G-D and 2D parts of the spectra shown in Fig. 3 separately.

## 6. One-phonon Raman spectra of $sp^2$ amorphous carbons

Figure 6 presents a collection of partitioned G-D spectra of the studied samples. Taking into account the fact that the G-D spectra in the language of C=C stretchings tell us about the structure of the carbon backbone of graphenous molecules, once the BSUs of the studied ACs, we will try to know this story. The first thing that is immediately obvious is the convincing testimony of the BSUs honeycombs structure of all amorphics. This feature strongly support the fragmented character of $sp^2$ ACs. The second thing, equally obvious, is that the pattern of the amorphic honeycombs, common overall, is noticeably different in details revealing by the difference in the spectra band-shapes. The main goal of the current analysis is to find out which namely information related to atomic structure of the studied BSUs is hidden in these details.

The broadbandness of the obtained RSs is quite expected and is caused by two natural reasons attributed to internal and external dynamic inhomogeneity of the relevant BSUs. The former is due to a large number of stretching modes accompanied with a significant dispersion of the corresponding frequencies while the second is caused by a principal non-stadardness of the BSUs atomic structure. The first reason is obviously due to the large number of carbon atoms in BSU molecules $\mathcal{N}$. Despite the fact, there is still an opportunity to divide $3\mathcal{N}$-6 molecular vibrations into a relatively small number of groups, which can be attributed to the characteristic modes originated from the modes of the pristine benzene molecule. As shown earlier [27], in a graphenous molecule with average lateral dimensions of ~ 2 nm and $\mathcal{N}$ ~ 200, approximately 200 vibrational modes can be attributed to C=C stretchings, filling the frequency range 1300–1600 cm$^{-1}$. The stretchings frequencies are determined by the C=C bond lengths, as a result of which the dispersions of bond lengths and vibrational frequencies are tightly connected to each other.

The external inhomogeneity of BSUs is resulted from the dependence of the C=C bond length distribution over the molecule honeycomb on the molecule shape and the configuration of terminating heteroatoms in the molecule circumference (see a detailed description of this link in [70, 71]). The smaller is molecule, the more is the influence of the two last facts on the C=C bond distribution. As shown [26], a fixed C:O:H atomic content can be implemented with a large number of molecular graphenous structures of both the same and different shape, which is the reason of the discussed BSUs non-standardness and unavoidable structural and dynamical inhomogeneity of the short-range order of $sp^2$ ACs. In its turn, the feature evidently promoted broadening of the resulting G and D bands.

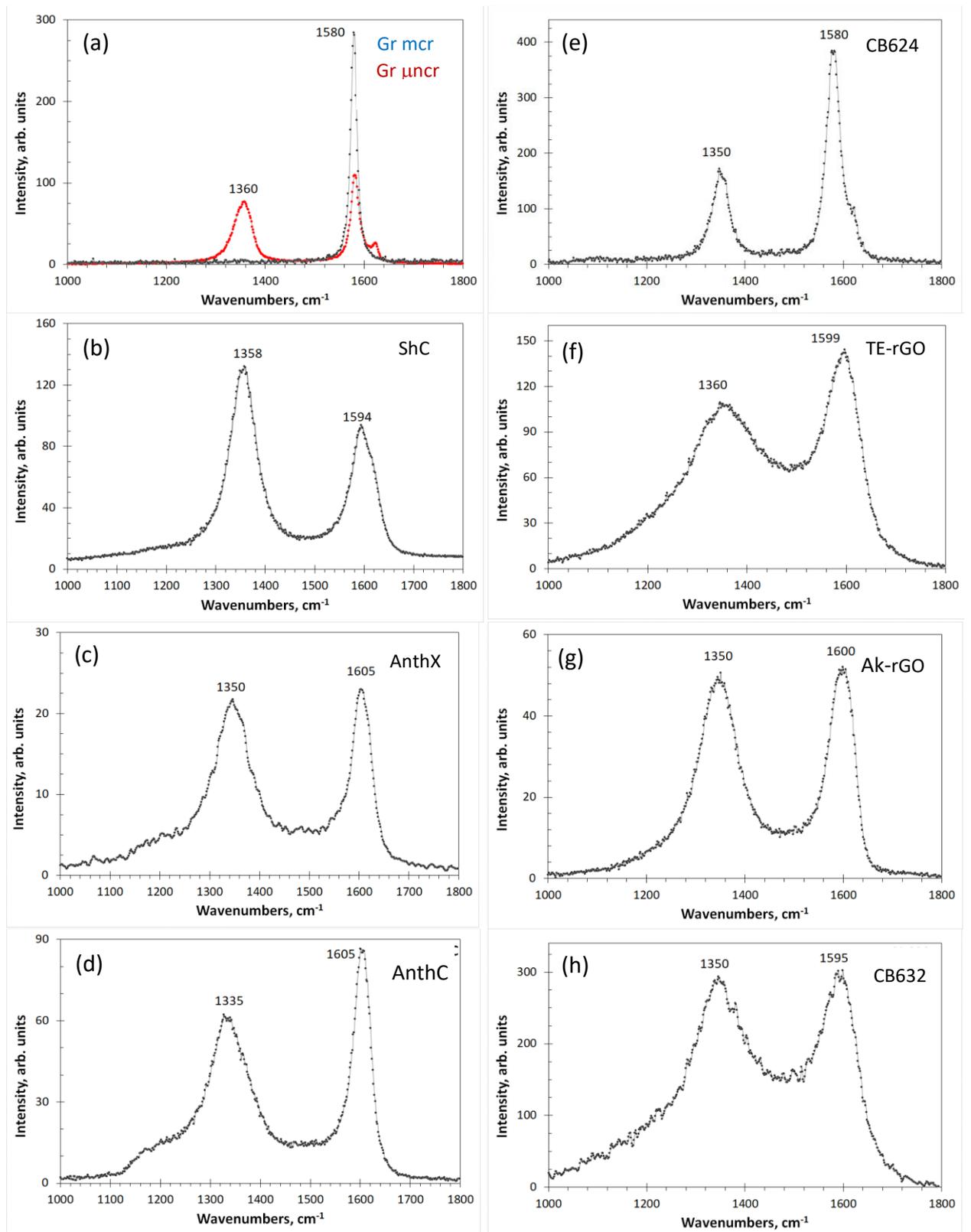

**Figure 6**. One-phonon Raman spectra of $sp^2$ amorphous carbons. Sample marking is the same as in Fig. 3.

The band-shape analysis of the spectra in Fig. 6 from this standpoint allows making the following conclusion.

(i) The largest BSU inhomogeneity is characteristic for technical graphene TE-rGO (Fig. 6f) while the smallest should be attributed to carbon black CB624 (Fig. 6e). BSUs of both amorphics are commensurate in lateral dimension with size ~ 20 nm (Table 1). However, when, as seen in the figures, big lateral size of CB624 BSU promotes an evident similarity of the spectra of this amorphic and μncr graphite (the issue of a particular interest will be discussed in details in Section 7), the difference between the spectra of technical graphene and micronanostructured graphite is the biggest. Apparently, the feature explanation should be associated with extremely complex chemical structure of the TE-rGO BSU circumference [27] that greatly disturbs C=C bond length distribution resulting in the large dispersion of stretchings frequencies.

A comparative analysis of G-D spectra of technical graphenes in Figs. 6f and g makes the last conclusion not so unambiguous. As seen, the two spectra differ drastically although the amorphics BSUs are commensurate by size and similar in complexity of the chemical composition on their circumference [27]. Nevertheless, G-D spectrum of Ak-rGO reveals much smaller inhomogeneity in comparison with TE-rGO and shows a remarkable similarity with the spectrum of, say, AnthX, in Fig. 6c, the BSU size of which is one order of magnitude less (see Table 1) while the circumference chemical composition much simpler [27].

(ii) G-D spectrum of carbon black CB632 in Fig. 6h differs from both the spectra of natural amorphics with commensurate BSUs and another carbon black CB624, BSU of which differs in size by more than one order of magnitude, demonstrating more significant dynamical inhomogeneity. As seen from the data presented for this amorphic in Fig. 7, changes in the chemical composition of the circumference is not enough to explain so big difference with respect to natural species. The largest that could be expected should be similar to changes in the spectrum of anthracite. In fact, the spectrum broadening turned out to be much stronger, which makes us look for other reasons for the feature such as the turbostratic character of BSU packing in stacks that is characteristic for CB632 [27] thus manifesting the effect of neighboring layer on the length distribution of C=C bonds in each individual layer.

(iii) G-D spectra of natural amorphics in Figs. 6b-d, which are characterized by BSUs of practically the same size (see Table 1), demonstrate a considerable dynamic inhomogeneity attributed to the C=C bond length distribution as well as the difference in the honeycomb packing related to their BSUs. The latter convincingly confirms the conclusion previously made in the analysis of the amorphics DRIFT spectra [27]. Figure 7 presents a set of BSU models attributed to the studied amorphics based on a joint analysis of DRIFT and XPS spectra. They all have the same carbon core consisting of 66 carbon atoms, only one (III), two (IV), and four (II) of which are substituted by oxygen atoms. The set is complemented with the C=C bond length distribution related to each model. As seen in the figure, the distribution remarkably changes in response to varying compositions of the models circumference thus evidencing a different assortment of both bonds and vibrational frequencies of each amorphic. Certainly, these changes should influence the band-shape of the final spectra.

(iv) The analysis performed above convincingly shows that the difference in the bond length distributions is obviously expected to be the main reason of the complicated broadband structure of the observed G-D spectrum. As seen in Fig. 7, all the distributions have a clearly seen frequency-stepped character. It is natural to expect that each step of these distributions in the real Raman spectrum should be associated with a relatively narrow band, because of which the observed spectrum represents a convolution of such bands. Obviously, the number and width of the bands depend on the step grouping of the bonds, shown for the studied models in Fig. 8a. According to the vibrational dynamics of molecules, grouping of valence bond lengths $\{l_i\}$ naturally transforms in that one of force constants $\{f_i\}$, which depend on the relevant lengths. Following Badger's rule [72], this dependence for C=C stretching vibrations takes a form $f_i = 3.0(l_i + 0.61)^{-3}$ [73] for constant and length units in $10^2$ Nm$^{-1}$ and $10^{-10}$ m, respectively. According to this relation, changing the length from 1.5 Å to 1.3 Å in Fig. 8a corresponds to frequencies growing from 1300 cm$^{-1}$ to 1700 cm$^{-1}$, which fully covers the spectral area of broadband G-D spectra observed experimentally. Therefore, grouping of C=C bonds in the studied amorphics causes a similar response of C=C stretchings frequencies thus laying the foundation of 'multiband' origin of their observed broadband G-D RSs.

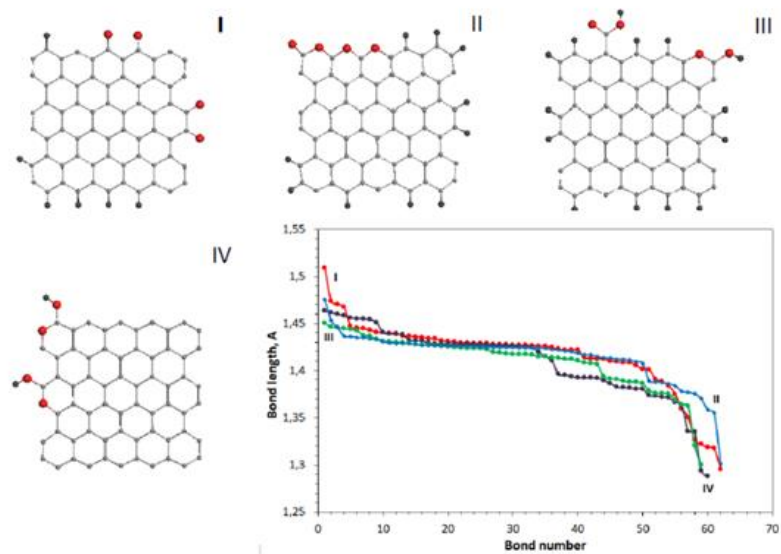

**Figure 7**. Equilibrium structure of BSU models related to shungite carbon (I), anthraxolite (II), anthracite (III), and carbon black CB632 (IV). Distribution of C=C bond lengths of the model graphene cores. UHF AM1 calculations.

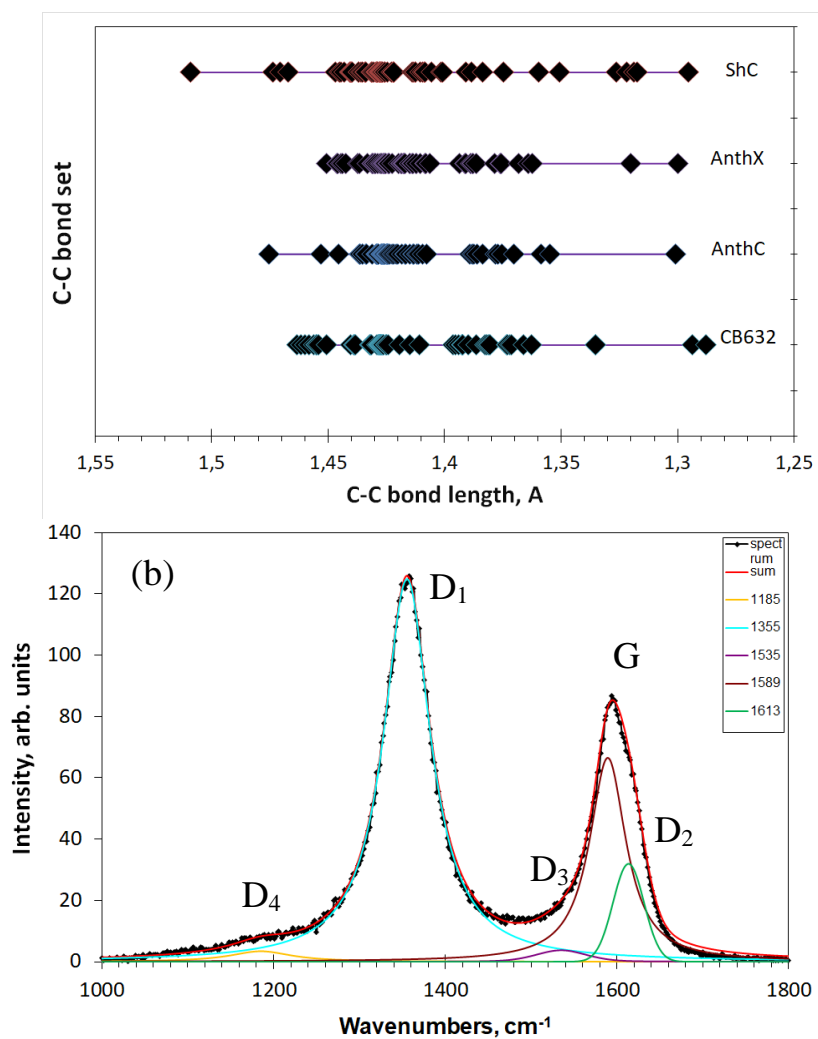

**Figure 8.** C=C bond sets of the model graphene cores (top) and deconvoluted Raman spectrum of shungite carbon (see text).

(v) Intuitively, the multiband character of the G-D spectra was taken into account and adopted by spectroscopists, starting with the first studies of Raman scattering in amorphous carbon and graphites (see, for example, [32, 66] and references therein). Standard programs of spectra deconvolution, such as LabSpec 5.39 in the current study, were widely used for experimental spectra decomposition. A typical example of such treatment is presented in Fig. 8b. The set of D1-D4 bands complemented with G band presents the usual basis for decomposing. Quite narrow spectral region and governing role of D1 and G bands provide a rather small dispersion in the maximum positions and FWHMs of D2-D4 bands thus giving a possibility to use the D1-D4 and G bands features when comparing RSs of different samples. Until now, it has been a comfortable formal language only facilitating the spectra description, while any changing G-D spectrum shape directly exhibits the reconstruction of the C=C sets of honeycomb structures of the samples BSUs.

(vi) Concluding the analysis of G-D spectra, it should note on the connection of widely used parameter $I_D/I_G$ with the real size of graphenous fragments. The advantage of this study, which allows us considering directly the relationship of this parameter with the size of the graphenous fragments, is the availability of independent data on the size of the studied BSUs obtained in previous studies [26, 27] and presented in Table 1. Basing on these data, we come to unexpected results. So, for natural amorphics (spectra in Figs. 6b-d), the BSU sizes of which are approximately the same, the $I_D/I_G$ doubles. The parameter for CB632 (Fig. 6h) of practically the same lateral size is close to that for AnthX, but differs from both ShC and AnthC. The same can be said about the parameters for spectra CB624, TE-rGO, and Ak-rGO (Figs. 6e-g) with similar size parameters. Thus, the parameter $I_D/I_G$ can hardly be considered as an indicator of the size of graphenous fragments. The same conclusion was reached by the Italian researchers on the analysis of the G-D spectra of PAHs [39], which, in itself, is quite obvious, since the main Ex. (1) determining the intensity of the G and D bands does not contain parameters indicating a direct addressing to the molecule size.

## 7. Two-phonon Raman spectra of *sp²* amorphous carbons

Second-order Raman spectra of graphene materials have a long and rich history. Initially discovered in different graphite materials in the form of doublet of bands at ~ 2720 cm$^{-1}$ (strong) and 3248 cm$^{-1}$ (very weak) [2, 75], designated as G '(then 2D) and 2G, respectively, this region of the spectrum demonstrating interesting properties has since been actively analyzed [76-79] (references are representative rather than exhaustive). Referring the reader to an excellent review of the state of art in this area [80], we will focus only on two features of the 2D band, which are important for further discussion. (i) There are no bands in the one-phonon spectrum of ideal graphite [75] and graphene [77] crystals, whose overtones or composite combinations can be assigned as the 2D band. Only in graphite and/or graphene materials, which are obviously devoid of an ideal structure, the source of the overtone is interconnected with the band D. This fact is tightly connected with the origin of the two-phonon spectrum of graphenous materials. According to the fundamentals of Raman scattering [60], anharmonic dependence on normal vibrations of both vibrational and electronic energies (so called mechanic and electric anharmonicities) leads to the appearance of IR and Raman vibrational spectra beyond one-phonon. Mechanic anharmonicity is described with the third derivatives of potential energy while electric one results from the second derivatives of the object polarizability in Ex.(1). The two features influence the band-shape of the two-phonon spectra differently. Thus, the inconsistency of the band-shapes of one- and two-phonon spectra found in the case of RS of graphenous crystals convincingly evidences a predominant role of the electric anharmonicity (*el*-anharmonicity below).

(ii) The second peculiarity of the 2D band concerns the surprising variability of its intensity with respect to the G band. So, in graphite crystal, the ratio of the total intensities $I_{2D}/I_G$ is ~ 1; in graphene crystal, it is equal to ~ 6 [78]; and in graphene whiskers it exceeds 13 [77]. Such large values and sharp fluctuations in the intensity of 2D bands, as well as the observation in graphite whiskers with an ideal crystal structure of a broad high order Raman spectrum located in the high-frequency region up to 7000 cm$^{-1}$, indicate an exceptionally big role for *el*-anharmonicity in this case. The authors are not aware of other examples of such a striking effect. Apparently, this property should be added to the treasure-box of graphenous materials uniqueness.

Despite the huge number of publications concerning RS of graphenous materials and the 2D band, in particular, the relationship of the 2D band with *el*-anharmonicity has not been considered. At the same time, modern vibrational spectroscopy attaches great importance to both mechanical and electrical anharmonicity, the participation of which provides good agreement between the experimental and calculated spectral data without fitting parameters [81–83]. Currently, this new approach can be applied to mid-size molecules such as thiophene or naphthalene, but the work on computational modules continues, so that graphenous molecules can be apparently considered in the near future. Nevertheless, the already obtained data on the quantitative accounting of anharmonicity in small molecules allow judging that the above-described behavior of the 2D band intensity in ideal crystals at a qualitative level is quite expected, if to assume the anharmonic behavior of the vibrational and electronic spectra of graphene-like structures to be peculiar. We dare to suggest that the highly delocalized character of the electron density of the graphene and/or graphite crystal [36] contributes to such a pronounced anharmonicity. Moreover, the role of this feature in mechanical effect is evidently not direct. In contrast, the second derivatives of polarizability are directly determined by the state of the electronic system, which, possibly, determines the special role of *el*-anharmonicity in graphene.

Delocalization of electron density is characteristic not only of graphite and/or graphene crystal, but also of graphenous molecules [84]. If the assumption stated above is true, then a second-order spectrum should also be observed in RSs of such molecules, of PHAs, for example, largely discussed in the previous section. It should be expected that the intensity of this spectrum will be lower than the total intensity of the G and D bands, since experimental studies have shown that with the destruction or "molecularization" of graphite [75, 76] and graphene [5-7, 78, 80], the intensity of the two-phonon spectrum decreases and its band-shape broadens. It is this kind of spectrum that was recorded in the case of two PAH molecules [42]. Looking at the spectra from this standpoint, we see that, actually, both spectra consist of the doublet of well-defined and narrow bands G and D located at 1603 and 1316 cm$^{-1}$, accompanied by an intense broad band with weakly expressed maxima at ~2610 , ~2835, and ~2910 cm$^{-1}$. The first and last frequency markers are in good agreement with the frequencies of 2D and D + G bands. The appearance of the second maximum is obviously associated with the combination bands whose sources in the one-phonon spectrum have yet to be determined thus supporting the *el*-anharmonicity origin of the spectra. In general, the band-shape of RS of both molecules is similar to that one's of the studied ASs shown in Fig. 3. Consider these spectra now in more details.

Presented in Fig. 9 is a collection of two-phonon RSs of the studied $sp^2$ ACs. The first feature of the spectra is that a characteristic four-component structure, clearly distinguished in the spectra of two graphites (Fig. 9a) and carbon black CB624 (Fig. 9e), can be traced in the spectra of all other ACs. Three frequency markers at ~2700, ~2940, and ~3200 cm$^{-1}$, convincingly attributed to 2D, D+G, and 2G combinations, are steadily observed in all spectra. The fourth marker at 2440 cm$^{-1}$, whose assignment remains unclear, is hidden in the law-frequency tails of other spectra. Important to note that the marker involves at least one fundamental mode with frequency less than that of the lowest D mode and that modes of this kind have never been observed in the one-phonon spectra. This feature is one more argument in favor of the *el*-anharmonic origin of the second-order RS of graphenous materials.

Among the studied amorphics, the spectrum of black carbon CB624 occupies a central place. Thus, its close similarity with the spectrum of μncr graphite indicates that, despite the obvious difference in size of the short-range order region in the graphite and amorphic BSU, in both cases we are dealing with well-ordered structures making it possible to consider the spectra band-shape in the quasi-particle phonon approximation (widely reviewed in [79, 80]). As known from the solid-state physics, quasi-particle description, based on the conservation of translational symmetry, is applicable with respect to size-restricted bodies when the size exceeds a critical value characteristic to the particle under consideration. In the case of phonons, the phonon free path determines this critical size (see the discussion of the issue for amorphics with molecular structure in [53, 84-86]). The crystal-like band-shape of the RS of carbon black CB624 under consideration evidences that the free path of optical phonons in graphene crystal is ≤ 14.6 nm, as follows from Table 1. This conclusion is fully supported by the behavior of one-phonon spectra of these solids shown in Fig. 6.

On the other hand, comparing the spectrum of CB624 with those of natural amorphics in Fig. 9b-d shows that despite a decrease of BSUs size of the latter by more than an order of magnitude, the short-range ordering in these representatives significantly preserved. It looks like to be more regular for shungite carbon and of a lesser extent for anthraxolite and anthracite. However, there might be another reason for the difference in the spectra band-shape. The lack of both translational and point symmetry makes it possible to deal with *el*-anharmonicity concerning a large pool of the molecules C=C stretchings that cover a large frequency region. So far the theory of anharmonic RS [81-83] has not offered general regularities that govern the selection of particular modes for two-phonon spectra. Nevertheless, it is evident that this concerns a group of modes and when the pool is large, the group can be rather populated, which, in its turn, leads to a significant broadening of the two-phonon band-shape. Evidently, vibrational forms of the modes lay the foundation for the predominance in selection. It can be expected that changing vibrational forms will obviously lead to the change of the two-phonon band-shape. BSUs of shungite, anthraxolite and anthracite belong to different families of graphenous molecules, C=C stretchings of which possess different vibrational form, at least partially.

And finally, a comparison of the CB624 spectrum with the spectra of technical graphenes in Fig. 9f-g reveals how strikingly can differ the short-range order of graphenous amorphics, characterized by the commensurate sizes of BUSs differing in the chemical composition of their circumferences. The issue perfectly supports the discussion concerning the role of vibrational forms presented above. Thus, the presented collection of spectra reflects a wide range of possible structural formations underlying amorphous carbons, the types and difference between which can be revealed using the characteristic features of the two-phonon Raman scattering spectra. It should be noted that the two-phonon spectra turn out to be more sensitive in this case than one-phonon ones.

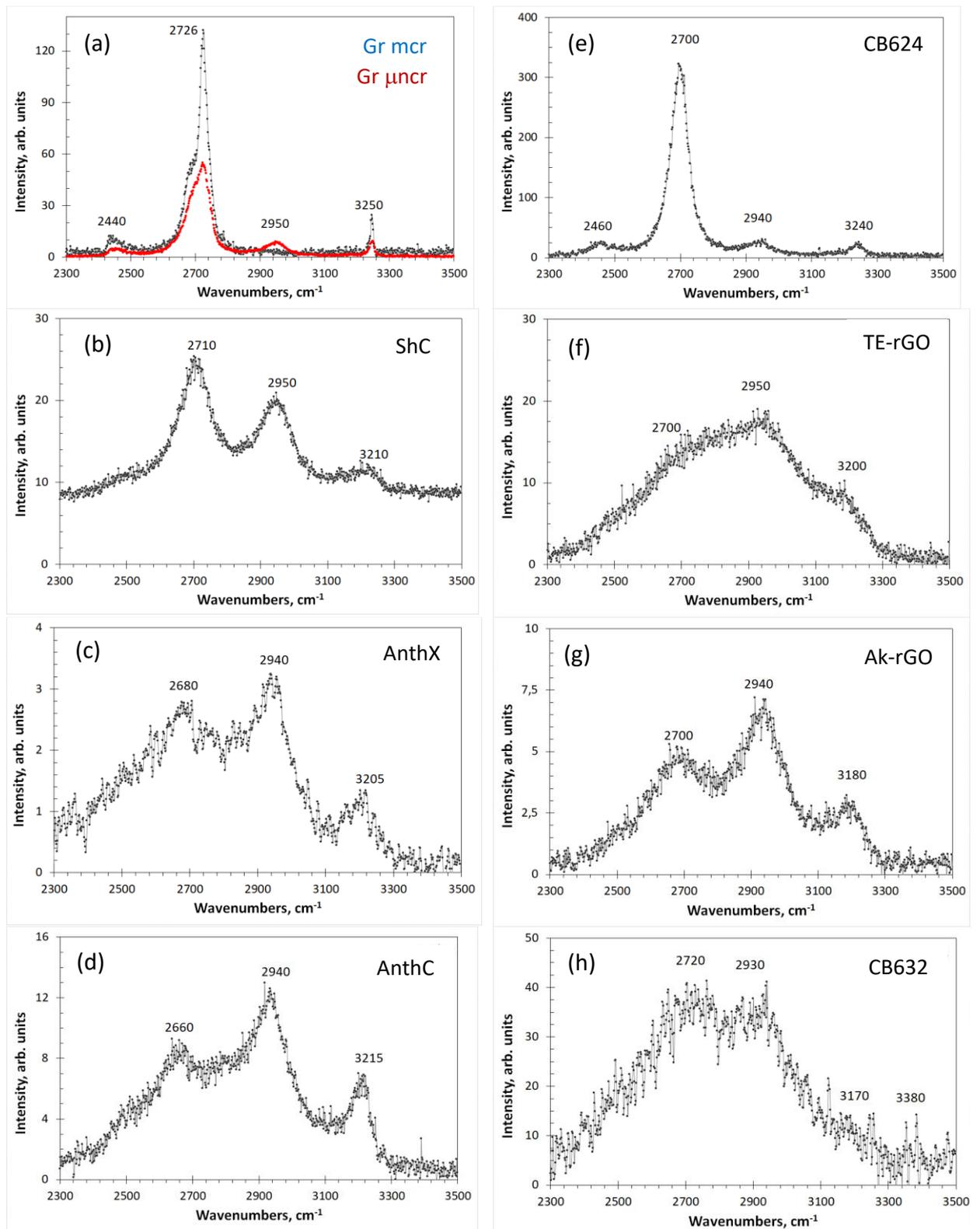

**Figure 9**. Two-phonon Raman spectra of *sp²* amorphous carbons. Sample marking is the same as in Fig. 3.

Concluding the analysis, one more thing should be mentioned. The frequency range of the studied two-phonon spectra coincides with the region of characteristic group frequencies (GFs) [74, 87], related to C-H stretching vibrations. Since, as seen in Table 2, all the studied amorphics, besides carbon blacks, are significantly hydrogenated, we tried to find the evidence of the hydrogen presence. However, none of the features related to the spectra observed can be attributed to characteristic GFs, such as 3050 cm$^{-1}$ related to methine groups [67] of natural amorphics as well as 2870-2970 cm$^{-1}$ and 2920-2980 cm$^{-1}$ of methylene and methyl groups [67] of technical graphenes TE-rGO and Ak-rGO, respectively. If hidden inside the broad bands, they might be revealed by a particular technique to be developed.

## 8. Conclusion

Long-term and numerous studies of amorphous solids have led researchers to the idea that a confident interpretation of their Raman spectra is inseparable from the consideration of the nature and type of the solids amorphization. In the current work, these two aspects were considered together with respect to *sp$^2$* ACs. Previous structural and analytical studies of these solids were analyzed from this viewpoint, due to which it was established that the they can be attributed to molecular amorphics of a new type of amorphism, which can be called *enforced fragmentation*. Chemical reactions occurred at the fragment edges are suggested to be one of the most important causes of this particular disordering. The fragments are stable graphenous molecules, once the basic structural units (BSUs) of ACs of various origins. The weak wdW interaction between the BUSs makes them the main defendants for the numerous properties of solids, including their IR absorption and Raman scattering. In this connection, Raman spectra of *sp$^2$* ACs are considered at the molecular level, which is perfectly suitable to case in contrast with just formal and very ill-founded, although widely used, double resonant Raman scattering approach. The molecular approximation is confirmed by the detail similarity of the spectra of studied solids and 2D PAHs studied earlier. The similarity itself evidences a governing role of the 2D honeycomb structure of both BSUs and PAHs due to which a standard G-D image of the spectra remains until the structure is fully destroyed.

The application of molecular approximation to the analysis of one-phonon spectra of the studied *sp$^2$* ACs makes it possible to explain the broadband shape of the G-D spectra as the evidence of a considerable dispersion of the molecules C=C bonds. Additionally, it is possible to associate individual differences in the spectra with different degrees of the dispersion, caused by the influence of chemical action, deformation, etc. It is this last circumstance that connects details of the Raman spectra of the AC BSUs in question with the history of the origin and production of the studied solids.

In the current study, the molecular approximation was first applied to the analysis of the two-phonon spectrum of *sp$^2$* ACs. Based on the fundamentals of the Raman scattering by molecules, explaining its origin by inducing a dipole moment, which is determined by the dependence of the molecule polarizability on normal coordinates, the features of the two-phonon spectrum of graphenous molecules are explained by a particular role of electrical anharmonicity. A suggestion concerning the special role of this effect in the carbon honeycomb structures addresses a high degree of the electron density delocalization.

Despite a perfect suitability of molecular approach to the interpretation of both one- and two-phonon spectra of the studied amorphics, it was possible to trace the transition from the molecular to the quasi-particle phonon approach when the lateral size of the ACs BSUs reaches

14.6 nm. The value can be used as an evaluation of a mean free path of optical phonons of graphite and/or graphene crystals originated from C-C stretchings.

In conclusion, it should be noted that the molecular approach makes it possible to carry out a coherent consideration of both one- and two-phonon spectra of $sp^2$ ACs, closely linking spectral individual characteristics with the structural and chemical features of the species BSUs. At the same time, once caused by the vibrations of carbon atoms, Raman spectra of these solids depend on chemical impurities indirectly through the influence of the latter on the C=C bond distribution, which makes the spectral technique non-applicable for a direct testing of the full chemical composition.

**Acknowledgements**

The authors are thankful to S.I. Isaenko for assistance in spectra recording and B.A. Makeev for XRD measurements. The study was performed as a part of research topics of the Institute of Geology of Komi Science Center of the Ural Branch of RAS (Grant No. AAAA-A17-117121270036-7). The publication has been prepared with the support of the "RUDN University Program 5-100".

**Statement about conflicts**

There are no conflicts of interests between the authors